\documentclass[12pt,preprint]{aastex}

\newcommand{\eqb}{\begin{equation}}
\newcommand{\eqe}{\end{equation}}

\begin{document}
\title{Induced scattering of short radio pulses}
\author{Yuri Lyubarsky}
\affil{Department of Physics, Ben-Gurion University, Beer-Sheva,
Israel}

\begin{abstract}
Effect of the induced Compton and Raman scattering on short,
bright radio pulses is investigated. It is shown that when a
single pulse propagates through the scattering medium, the
effective optical depth is determined by the duration of the pulse
but not by the scale of the medium. The induced scattering could
hinder propagation of the radio pulse only if close enough to the
source a dense enough plasma is presented. The induced scattering
within the relativistically moving source places lower limits on
the Lorentz factor of the source. The results are applied to the
recently discovered short extragalactic radio pulse.
\end{abstract}

\section{Introduction}
Induced scattering could significantly affect radiation from
sources with high brightness temperatures. The induced Compton
scattering may be relevant in pulsars (Wilson \& Rees 1978;
Lyubarskii \& Petrova 1996, Petrova 2004a,b, 2007a,b), masers
(Zeldovich, Levich \& Sunyaev 1972; Montes 1977) and radio loud
active galactic nuclei (Sunyaev 1971; Coppi, Blandford \& Rees
1993; Sincell \& Coppi 1996). The induced Raman scattering is
considered as the most plausible mechanism of eclipses in
binary pulsars (Eichler 1991; Gedalin \& Eichler 1993; Thompson
et al. 1994; Luo \& Melrose 1995) and was also invoked to place
constraints on the models of pulsars (Lyutikov 1998; Luo \&
Melrose 2006) and models of intraday variability in compact
extragalactic sources (Levinson \& Blandford 1995).

Macquart (2007) used the induced Compton and Raman scattering
in order to place limits on the observability of the prompt
radio emission predicted (Usov \& Katz 2000; Sagiv \& Waxman
2002; Moortgat \& Kuijpers 2005) to emanate from gamma-ray
bursts. Recent discovery of an enigmatic  short extragalactic
radio pulse (Lorimer et al 2007) demonstrates that very high
brightness temperature transients do exist in nature. In this
paper, we address the induced scattering of short bright radio
pulses. First we study the induced Compton and Raman scattering
in the plasma surrounding the source. The central point is that
due to the non-linear character of the process, the effective
optical depth is determined  not by the scale of the scattered
medium but by the width of the pulse provided that the pulse is
short in the sense that duration of the pulse is less than the
light travel time in the scattered medium. For this reason,
short enough pulses could propagate through the interstellar
medium, contrary to Macquart's claim. The induced scattering
could hinder propagation of a high brightness temperature pulse
only close enough to the source if the density of the ambient
plasma is large enough; here we find the corresponding
observability conditions. We also address the induced
scattering within the relativistically moving source and show
that transparency of the source implies a lower limit on the
Lorentz factor of the source. We apply the general results to
the short extragalactic radio pulse discovered by Lorimer et
al. (2007).

\section{Induced Compton scattering}
The kinetic equation for the induced Compton scattering in the
non-relativistic plasma is written as (e.g. Wilson 1982)
 \eqb
\frac{\partial n(\nu,\mathbf{\Omega})}{\partial t}+
c(\mathbf{\Omega\cdot\nabla})n(\nu,\mathbf{\Omega})=
\frac{3\sigma_T}{8\pi}N \frac{h}{m_ec}n(\nu,\mathbf{\Omega})
 \int(\mathbf{e\cdot e}_1)^2
(1-\mathbf{\Omega\cdot\Omega}_1)\frac{\partial
\nu^2n(\nu,\mathbf{\Omega}_1)}{\partial\nu}d\mathbf{\Omega}_1;
  \label{kinComp}\eqe
where $n(\nu,\mathbf{\Omega})$ is the photon occupation number of
a beam in the direction $\mathbf{\Omega}$, $N$ the electron number
density, $\mathbf{e}$ the polarization vector. The induced
scattering rate is proportional to the number of photons already
available in the final state therefore  the scattering initially
occurs within the primary emission beam where the radiation
density is high. However, when the primary beam is narrow, as is
anyway the case at large distance from the source, the recoil
factor $1-\mathbf{\Omega\cdot\Omega}_1$ makes the scattering
within the beam inefficient; then the scattering outside the beam
dominates because according to Eq.(\ref{kinComp}), even weak
isotropic background radiation (created, e.g., by spontaneous
scattering) grows exponentially so that the energy of the
scattered radiation becomes eventually comparable with the energy
density in the primary beam.

In this and the next sections, we study the induced scattering
outside of the source therefore we can assume that the
scattering angle is larger than the small angle subtended by
the primary radiation. In this case, the occupation number of
the scattered photons varies according to the equation
 \eqb \frac
1n\frac{dn}{dt}=\frac{3\sigma_T}{8\pi} \frac{cN}{m_e}
(\mathbf{e\cdot e}_1)^2
(1-\cos\theta)\frac{\partial}{\partial\nu}\left(\frac
F{\nu}\right);
 \label{kinComp1} \eqe
where $F=c^{-2}h\int\nu^3nd\mathbf{\Omega}$ is the local radio
flux density of the primary radiation, $\theta$ the scattering
angle. Solution to this equation is written as
 \eqb
n=n_0\exp{\tau_C};
 \eqe
where $n_0$ is the background photon density, $\tau_C$ the
effective optical depth determined by the integral along the
scattered ray,
$\mathbf{r}=\mathbf{r}_0+c\mathbf{\Omega}(t-t_0)$, as
 \eqb
\tau_C=\int \frac{3\sigma_T}{8\pi} \frac{cN}{m_e}
(\mathbf{e\cdot e}_1)^2
(1-\cos\theta)\frac{\partial}{\partial\nu}\left(\frac
F{\nu}\right) dt.
 \label{tau}\eqe
The intensity of the scattered radiation increases
exponentially provided the photon spectrum of the primary beam,
$F/\nu$, has a positive slope. Therefore the induced scattering
is the most efficient just below the spectral maximum. If the
radiation with a decreasing spectrum is detected, one can find
the observability condition substituting the frequency
derivative in Eq.(\ref{tau}) by $F/\nu^2$ at the observed
frequency because the stimulated scattering rate thus estimated
is lower than that near the spectral maximum. As the brightness
temperature of the primary beam many orders of magnitude
exceeds the brightness temperature of the background radiation,
the fraction of the scattered photons remains small until
$\tau_C$ reaches a few dozens. As a simple criterion for the
observability of the primary radiation (the condition that the
induced scattering does not affect the primary radiation), one
can use the condition $\tau_C<10$.

In order to check this condition, one can substitute  the
undisturbed primary flux into Eq.(\ref{tau}). Let a radio pulse of
the duration $\Delta t$ propagate radially from the source; then
the primary flux can be presented in the form
 \eqb
F=\left(\frac{D}r\right)^2F_{\rm obs}
\Theta\left(\frac{ct-r}{c\Delta t}\right);
 \label{flux}\eqe
where $D$ is the distance to the source, $r$ the distance from
the source to the scattering point, the function $\Theta(x)$
describes the shape of the pulse.  Below we adopt the simplest
rectangular form: $\Theta(x)=1$ at $0<x<1$ and $\Theta(x)=0$
otherwise.  Note that we can ignore the transverse structure of
the pulse because the most efficient is the backscattering so
that the scattered ray interacts only with the radiation
emitted in the same direction.

Note also that even though Eq.(\ref{flux}) assumes that the
pulse structure is attributed to the intrinsic time variation
of the source, the same structure arises if pulsed radiation is
generated by a narrow beam sweeping across the observer. In
this case, the radiation field has a shape
$\Theta\left[(ct-r-r_0\varphi)/c\Delta t\right]$, which is
reduced to Eq.(\ref{flux}) at $\varphi=\it const$. However, one
should take into account that in this paper, we assume that the
pulse is {\it single} in the sense that the distance between
pulses is larger than the scale of the scattering medium. If
this condition is not fulfilled, the scattered ray could pass
through a few pulses and then the induced scattering occurs as
in the steady radiation field with the intensity equal to the
average intensity of the source. Therefore the results of this
paper should be applied only to true single events like radio
emission from gamma-ray bursts or giant pulses from pulsars,
which are rare enough to be considered as isolated phenomena.

Let a seed ray be launched in the point $r_0$ at the time
$t_0=cr_0$ just when the pulse reached this point.  Due to
induced scattering of the photons from the pulse, the intensity
of the ray grows exponentially while the ray remains within the
zone illuminated by the pulse. Below we assume that the pulse
is narrow enough, $c\Delta t\ll r_0$. In order to find the
amplification factor of the seed ray, one should find the
effective optical depth (\ref{tau}), which could be presented
as
 \eqb
\tau_C=\frac{3\sigma_T}{8\pi} \frac{cNF_{\rm
obs}}{m_e\nu^2}\left(\frac{D}{r_0}\right)^2Z
 \label{tau1}\eqe
where
 \eqb
Z=\int (1-\cos\theta)\left(\frac{r_0}r\right)^2
\Theta\left(\frac{ct-r}{c\Delta t}\right)dt
 \label{int}\eqe
is the integral along the ray. For the estimates, we take
$\mathbf{e\cdot e}_1=1$. Let the ray be directed at the angle
$\theta_0$ to the radial direction at the initial point. Then the
scattering angle, $\theta$, and the distance from the source, $r$,
at the time $t$ could be found from the laws of sines and cosines
for the triangle in Fig. 1
 \eqb
\frac{c(t-t_0)}{\sin(\theta_0-\theta)}=\frac{r_0}{\sin \theta};
 \label{sin}\eqe \eqb
r^2=r_0^2+c^2(t-t_0)^2+2r_0c(t-t_0)\cos\theta_0.
 \label{cos}\eqe
Eliminating $r$ and $t$, one can present the integral
(\ref{int}) as
 \eqb
Z=\frac{r_0}{c\sin\theta_0}\int_{\theta_{\rm
min}}^{\theta_0}(1-\cos\theta)\theta d\theta
 =\frac{r_0}c\frac{\theta_0-\theta_{\rm
min}-\sin\theta_0+\sin\theta_{\rm min}}{\sin\theta_0};
 \eqe
where $\theta_{\rm min}$ is determined from the condition that
the function $\Theta$ vanishes, $r=c(t-\Delta t)$. This
condition, together with Eqs.(\ref{sin}) and (\ref{cos}),
yields the equation for $\theta_{\rm min}$
 \eqb
\frac{\tan\theta_0}{\tan\theta_{\rm min}}-1=\frac{\Delta t}
{2t_0\cos\theta_0} \frac{2-\Delta t}{t_0(1-\cos\theta_0)-\Delta
t}.
 \eqe
Taking into account that $\theta_0-\theta_{\rm min}\ll 1$ at
$\Delta t\ll t_0$, one gets
 \eqb
\theta_{\rm min}=\left\{\begin{array}{ll}\theta_0-\frac{c\Delta
t\sin\theta_0}{r_0(1-\cos\theta_0)};
& \theta_0>\sqrt{\frac{2c\Delta t}{r_0}}; \\
0 & \theta_0<\sqrt{\frac{2c\Delta t}{r_0}}.\end{array}\right.
 \eqe
If $\theta_0<\sqrt{2c\Delta t/r_0}$, the scattered ray remains
within the illuminated area till infinity therefore
$\theta_{\rm min}=0$ in this case. Finally one finds
 \eqb
Z=\left\{\begin{array}{ll} \Delta t\left(1-\frac{2c\Delta
t}{r_0\theta_0^2}+\frac{4c^2\Delta t^2}{4r_0^2\theta_0^4}\right);
&  \theta_0>\sqrt{\frac{2c\Delta t}{r_0}}; \\
\frac{r_0}{6c}\theta_0^2 & \theta_0<\sqrt{\frac{2c\Delta
t}{r_0}}.\end{array}\right.
 \eqe
One sees that the amplification factor is the same for all the
scattered rays launched at not too small angles,
$\theta_0\gg\sqrt{2c\Delta t/r_0}$. This is because decreasing
of the scattering rate with decreasing angle (due to the recoil
factor $1-\cos\theta$ in the scattering rate) is compensated by
increasing of the time the scattered ray spends within the
illuminated area. The rays launched at the angles
$\theta_0\lesssim\sqrt{2c\Delta t/r_0}$ spend within the
illuminated area the time $t-t_0\gtrsim r_0/c$; then the
amplification factor decreases because of decreasing of the
primary radiation density with the distance. Of course if the
amplification factor is large, it is the backscattered
radiation that takes the whole energy of the primary beam
because the backward scattering is the fastest.

Substituting $Z=\Delta t$ into Eq.(\ref{tau1}) one can now
estimate the effective optical depth to the induced scattering;
numerically one gets
 \eqb
\tau_C=0.24\frac{N_6\Delta t_{\rm s}F_{\rm obs, Jy}}{\nu_{\rm
GHz}^2}\left(\frac{D_{8}}{r_{-3}}\right)^2;
 \eqe
where $\Delta t_{\rm s}$, $F_{\rm obs, Jy}$ and $\nu_{\rm GHz}$
are measured in units shown in the index, $D=10^8D_8$ pc,
$N=10^6N_6$ cm$^{-3}$, $r_0=10^{-3}r_{-3}$ pc. One sees that
the induced scattering is negligible in the interstellar medium
however it could become significant in dense enough environment
close enough to the source. For example a massive star could be
a progenitor of the gamma-ray burst; then the emission
propagates through the relic stellar wind. In this case the
plasma density falls off as
 \eqb
N_6=0.03\frac{\dot{M}_{-5}}{V_3 r_{-3}^2};
 \label{wind}\eqe
where $\dot{M}=10^{-5}\dot{M}_{-5}$ $M_{\bigodot}\cdot$yr$^{-1}$
is the mass loss rate, $V=10^3V_3$ km$\cdot$s$^{-1}$ the wind
velocity. The condition $\tau_C<10$ places the lower limit on the
radius beyond which the radio pulse could propagate:
 \eqb
r_{-3}>0.16\left(\frac{D_{8}}{\nu_{\rm
GHz}}\right)^{1/2}\left(\frac{\Delta t_{\rm s}F_{\rm obs,
Jy}\dot{M}_{-5}}{V_3}\right)^{1/4}.
 \label{windC}\eqe

\section{Induced Raman scattering}

The high intensity radio beam could be scattered by emitting
Langmuir waves. The energy and momentum conservation in this
three-wave process require that
 \eqb
\nu_1=\nu+\nu_p;\qquad \mathbf{k}_1=\mathbf{k}+\mathbf{q};
 \label{cons}\eqe
where $\nu_p=\sqrt{e^2N/(\pi m_e)}$ is the plasma frequency,
$\mathbf{q}$ the wave vector of the plasma wave. In the case
$\nu\gg\nu_p$ one can neglect the frequency shift of the
scattered wave; then one finds
 \eqb
\mathbf{q}_{\pm}=\pm\frac{\omega}c
(\mathbf{\Omega}_1-\mathbf{\Omega}).
 \label{qu}\eqe
Here the sign $+$ is associated with a plasmon emitted by the
photon $\mathbf{k}_1$ and the sign $-$ with a plasmon emitted
by the photon $\mathbf{k}$. Because of Landau damping, only
plasmons with large enough phase velocities could survive; this
places a limit on the scattering angle (Thompson et al. 1994).
Namely, choosing the allowable range of the plasmon wavevectors
from the condition that the Landau damping time exceeds the
period of the plasma wave, $q\lambda_D<0.27$, where
$\lambda_D=\sqrt{k_BT/(4\pi e^2N)}$ is the Debye length, one
finds from Eq. (\ref{qu}) that the backscattering is possible
only if $\nu<\nu_L=90N_6^{1/2}T_6^{-1/2}$ MHz. In the case
$\nu\gg\nu_L$, the maximum angle of scattering is
 \eqb
\theta_{\rm max}=2\nu_L/\nu.
 \label{maxangle}\eqe

The kinetic equations for the occupation numbers of photons and
plasmons are written as (Thompson et al. 1994)
 \eqb
\frac{\partial n(\nu,\mathbf{\Omega})}{\partial
t}+c(\mathbf{\Omega\cdot\nabla})n(\nu,\mathbf{\Omega})
 \label{kinRamanph}\eqe
$$=\frac{3\sigma_T}{8\pi}N \frac{h\nu}{m_ec}\frac{\nu}{\nu_p}
\int (\mathbf{e\cdot e}_1)^2(1-\mathbf{\Omega\cdot\Omega}_1)
[n_{q}+n_{-q}]
[n(\nu,\mathbf{\Omega}_1)-n(\nu,\mathbf{\Omega})]d\mathbf{\Omega}_1;
 $$\eqb
\frac{\partial n_{\pm q}}{\partial t}+v_{\rm
g}((\mathbf{q}/q)\cdot\nabla)n_{\pm q}
  \label{kinRamanpl}\eqe
  $$=\frac{3\sigma_T}{8\pi}N
\frac{h\nu}{m_ec}\frac{\nu}{\nu_p}\int (\mathbf{e\cdot e}_1)^2
\{n(\nu,\mathbf{\Omega}_1)n(\nu,\mathbf{\Omega})\pm n_{\pm
q}[n(\nu,\mathbf{\Omega}_1)-n(\nu,\mathbf{\Omega})]\}d\mathbf{\Omega}_1
-2\kappa n_{\pm q};
 $$
where $n_q$ is the plasmon occupation number, $\kappa$ the
plasmon amplitude damping rate, $v_{\rm
g}=3q\lambda_D\sqrt{k_BT/m_e}$ the plasmon group velocity. The
last is small in the non-relativistic plasma therefore one can
neglect the spatial transfer of plasmons.

As in the previous section, we assume that the primary
radiation subtends the angle smaller than the scattering angle
(\ref{maxangle}); then the scattering occurs outside the
primary beam because the scattering within the beam is
suppressed by the factor $1-\mathbf{\Omega\cdot\Omega}_1$. As
in the previous section, we will find the observability
condition demanding that the amplification factor of a weak
background radiation due to the Raman scattering does not
become exponentially large. One should stress that the Raman
scattering does not necessary hinders propagation of the
radiation even if the effective optical depth is large because
the scattering angle (\ref{maxangle}) may be small. Then the
radiation beam just widens and a special analysis is necessary
in order to figure out how much parameters of the emerged
radiation are affected. An example of such an analysis is given
in sect. 5. Here we just find the effective optical depth to
the Raman scattering.

Assuming that the primary pulse has the form (\ref{flux}) and
that the intensity of the scattering radiation is small as
compared with the primary radiation, one reduces the kinetic
equations (\ref{kinRamanph}) and (\ref{kinRamanpl}) to the form
 \eqb
\frac{\partial n}{\partial t}+c\cos\theta\frac{\partial
n}{\partial r}
=S\left(\frac{r_0}r\right)^2(1-\cos\theta)(n_q+n_{-q})
\Theta\left(\frac{ct-r}{c\Delta t}\right);
 \label{kinRph}\eqe
 \eqb
\frac{\partial n_{\pm q}}{\partial t} =S\left[(n\pm n_{\pm
q})\left(\frac{r_0}r\right)^2-\alpha n_{\pm q}
\right]\Theta\left(\frac{ct-r}{c\Delta t}\right);
 \label{kinRpl}\eqe
where
 \eqb
S=\frac{3\sigma_T}{8\pi}\frac{cNF_0}{m_e\nu\nu_p}
\left(\frac{D}{r_0}\right)^2(\mathbf{e\cdot e}_1)^2 ;\qquad
\alpha=\frac{F_{\kappa}}{F_{\rm obs}};\qquad
F_{\kappa}=\frac{16\pi m_e\nu\nu_p}{3\sigma_TcN(\mathbf{e\cdot
e}_1)^2}\left(\frac{r_0}{D}\right)^2\kappa.
 \eqe
The plasmon decay rate due to electron-ion collisions is
$\kappa=0.032N_6T_6^{-3/2}\,{\rm s}^{-1}$; then
 \eqb
F_{\kappa}=2.2\times 10^{-3}\frac{N_6^{1/2}\nu_{\rm
GHz}}{T_6^{3/2}} \left(\frac{r_{-3}}{D_8}\right)^2\,\rm Jy.
 \label{Fkappa}\eqe
We assume that before the pulse arrives, some weak background
radiation preexists in the medium therefore the boundary
conditions may be written as
 \eqb
n\vert_{r=ct}=n_0;\qquad n_q\vert_{r=ct}=n_{q0}.
 \label{boundary}\eqe

The factor $(r_0/r)^2$ in the right-hand side of Eqs.
(\ref{kinRph}) and (\ref{kinRpl}) arises due to decreasing of
the primary radiation flux (\ref{flux}) with the distance. It
was shown in the previous section that if the scattering angle
is not too small, $\theta_0\gg\sqrt{2c\Delta t/r_0}$, the
scattered ray remains within the illuminated area only during
the time $t-t_0\ll r_0/c$; then the factor $(r_0/r)^2$ may be
substituted by unity. Taking into account that the maximal
scattering angle is given by Eq.(\ref{maxangle}), this
condition is written as
 \eqb
\frac{N_6r_{-3}}{T_6\Delta t_{\rm s}\nu^2_{\rm GHz}}\gg 6\times
10^{-4}.
 \label{condition}\eqe
In this case, Eqs. (\ref{kinRph}) and (\ref{kinRpl}) are easily
solved. Namely, transforming the variables
 \eqb
v=Sct;\qquad u=S(ct-r);
 \eqe
one comes to the set of equations
 \eqb \frac{\partial
n}{\partial v}+(1-\cos\theta)\frac{\partial n}{\partial
u}=(1-\cos\theta)(n_q+n_{-q})\Theta\left(\frac{u}{cS\Delta
t}\right);
 \eqe
 \eqb
\frac{\partial n_{\pm q}}{\partial v}+\frac{\partial n_{\pm
q}}{\partial u}=\left[n-(\alpha\mp 1)n_{\pm
q}\right]\Theta\left(\frac{u}{cS\Delta t}\right);
 \eqe
with the boundary conditions at the point $u=0$. As both
coefficients of the equations and the boundary conditions are
independent of $v$, the solution is also independent of $v$
therefore one finally gets a simple set of ordinary differential
equations at the segment $0<u<cS\Delta t$:
 \eqb
\frac{d n}{du}=n_q+n_{-q}; \qquad \frac{d n_{\pm
q}}{du}=n-(\alpha\mp 1)n_{\pm q}.
 \label{set}\eqe
The boundary conditions are: $n(0)=n_0$, $n_{\pm q}(0)=n_{q0}$.

Partial solutions to these equations have a form $\exp(su)$ where
$s$ obeys the characteristic equation
 \eqb
s^3+2\alpha s^2+(\alpha^2-3)s-2\alpha=0.
 \label{charact}\eqe
Simple solutions are found in the two limiting cases, namely
when one can neglect the decay of plasmons, $\alpha=0$, and
when the decay is strong, $\alpha\gg 1$ (these limits
correspond to the conditions that the primary radiation flux is
well above or well below the limiting flux (\ref{Fkappa}),
correspondingly). In the limit $\alpha=0$ the solution to
Eqs.(\ref{set}) is
 \eqb
n=\frac 13 [n_0+2n_0\cosh\sqrt{3}u
+2\sqrt{3}n_{q0}\sinh\sqrt{3}u];
 \label{solution1}\eqe\eqb
n_{\pm q}=\frac 13[\mp n_0+(3n_{q0}\pm n_{0})\cosh\sqrt{3}u
+\sqrt{3}(n_{0}\pm n_{q0})\sinh\sqrt{3}u].
 \eqe
In the limit $\alpha\gg 1$ the solution is
 \eqb
n=n_0\exp\left(\frac 2{\alpha}u\right);
 \label{solution2}\eqe\eqb
n_{\pm q}=\frac{n_0}{\alpha}\exp\left(\frac
2{\alpha}u\right)+\left(n_{q0}-\frac{n_0}{\alpha}\right)\exp[(\pm
1-\alpha)u].
 \eqe
The intensity of the scattered radiation grows until $u=u_{\rm
max}=S\Delta t$ so the effective optical depth to the Raman
scattering may be estimated as $\tau_R=S\Delta t$ in the case
$\alpha\ll 1$ and $\tau_R=S\Delta t/\alpha$ in the opposite
limit. Numerically one gets
 \eqb
\tau_R=29\frac{N_6^{1/2}\Delta t_{\rm s}F_{\rm obs}}{\nu_{\rm
GHz}}\left(\frac{D_8}{r_{-3}}\right)^2\left\{\begin{array}{ll}1;
& F_{\rm obs}\gg F_{\kappa}; \\
F_{\rm obs}/F_{\kappa}; & F_{\rm obs}\ll F_{\kappa}.
\end{array}\right.
 \label{tauR}\eqe

As in the case of the induced Compton scattering, the condition
for the Raman scattering to remain negligible may be written as
$\tau_R<10$. One can see again that the scattering in the
interstellar medium is negligible. Assuming that the emission
is generated within the stellar wind of the progenitor star
(see Eq.\ref{wind}), one obtains that the Raman scattering
could be neglected if the radio pulse was emitted at the
distance
 \eqb
r_{-3}>0.8D_8^{2/3}\left(\frac{\Delta t_{\rm s}F_{\rm
obs,Jy}}{\nu_{\rm GHz}}\right)^{1/3}
\left(\frac{\dot{M}_{-5}}{V_3}\right)^{1/6}\left\{\begin{array}{ll}1;
& F_{\rm obs}\gg F_{\kappa}; \\
(F_{\rm obs}/F_{\kappa})^{1/3}; & F_{\rm obs}\ll F_{\kappa}.
\end{array}\right.
\label{windR}\eqe
from the source.

This result was obtained under the condition (\ref{condition}),
i.e. if the scattering angle is not too small and the
interaction of the scattered ray with the primary pulse occurs
at the scale small enough that one can neglect decreasing of
the primary radiation flux with radius. Therefore we neglected
the factor $(r_0/r)^2$ in the right-hand side of
Eqs.(\ref{kinRph}) and (\ref{kinRpl}). In the opposite limit,
the scattered ray remains within the illuminated zone for a
long time however due to decreasing of the radiation flux, only
the region $r-r_0\sim r_0$ contributes to the effective optical
depth (cp. Eq.(13) and discussion after). As the solutions
(\ref{solution1}) and (\ref{solution2}) are valid at
$r-r_0<r_0$, one can find the amplification factor substituting
into these solutions $u_{\rm max}$ corresponding to the radius
$r=2r_0$. It follows from the scattering geometry (see Fig.1
and Eq.(\ref{cos})) that $u_{\rm max}=Sr_0\theta_0^2/(4c)$.
Substituting $\theta_0$ by the maximal scattering angle of
Eq.(\ref{maxangle}), one gets finally the estimate for the
effective optical depth at the condition opposite to that of
Eq.(\ref{condition})
 \eqb
\tau_R=2.4\times 10^4\frac{N_6^{3/2}F_{\rm
obs,Jy}D_8^2}{T_6\nu^3_{\rm
GHz}r_{-3}}\left\{\begin{array}{ll}1;
& F_{\rm obs}\gg F_{\kappa}; \\
F_{\rm obs}/F_{\kappa}; & F_{\rm obs}\ll F_{\kappa}.
\end{array}\right.
 \eqe
Note that within the range of applicability of this formula, it
gives the optical depth smaller than Eq.(\ref{tauR}).

\section{Induced scattering within a relativistic source}
If a high brightness temperature radio pulse is generated in a
relativistic source, one can restrict parameters of the source
considering stimulated emission within it. Let a radio pulse
come from a relativistically hot plasma moving with the Lorentz
factor $\Gamma$. In the comoving frame, the radiation could be
considered as isotropic; then the kinetic equation for the
induced Compton scattering could be written as (Melrose 1971)
 \eqb
\frac 1{n(\nu')}\frac{\partial n(\nu')}{\partial t'}=\frac
3{16}\sigma_T\frac{h\nu'}{m_ec}N'
 \int d\gamma\frac{\partial}{\partial\gamma'}\left(\frac{f(\gamma')}{\gamma'^2}\right)
\int_0^{\infty}\frac{d\nu'_1}{\nu'_1}\left(1-\frac{\nu'_1}{\nu'}\right)
g\left(\frac{\nu'_1}{\nu'}\right)n(\nu'_1);
 \label{eq_melrose}\eqe
where $f(\gamma')$ is the electron distribution function
normalized as $\int f(\gamma')d\gamma'=1$; the primed
quantities are measured in the comoving frame. The kernel $g$
is approximated as (correcting a typo in Melrose's paper)
 \eqb
g(x)=\left\{\begin{array}{ll}4x^2;
& (2\gamma')^{-2}\le x\le 1; \\
4x; & 1\le x\le 4\gamma'^2; \\
0; & {\rm otherwise}.\end{array}\right.
 \eqe
The right-hand side of Eq.(\ref{eq_melrose}) is the induced
scattering rate; it should be compared with the rate of photon
escape from the source, $c/l'$, where $l'$ is the characteristic
size of the emitting region. If the emitting plasma moves, as is
typically the case, radially from the origin, one should also take
into account that the plasma density decreases in the proper frame
with the rate $c\Gamma/r$. Then the condition that the induced
scattering does not affect the emerged radiation is written as
 \eqb
\frac 1{n(\nu')}\frac{\partial n(\nu')}{\partial t'}
<\max\left(\frac c{l'},\frac{c\Gamma}{r}\right).
 \label{ind_cond}\eqe

In order to estimate the stimulated scattering rate, let us
assume that the particle distribution is Maxwellian
 \eqb
f(\gamma)=\frac{\gamma^2}{2\gamma_T^3}\exp\left(-\frac{\gamma}{\gamma_T}\right)
 \eqe
and that the radiation spectrum has a form
 \eqb
I(\nu')=I_0\left\{\begin{array}{ll}(\nu'/\nu'_0)^a;
& \nu'<\nu'_0; \\
(\nu'/\nu'_0)^{-b}; & \nu'>\nu'_0;\end{array}\right.
 \eqe
where $I(\nu')=h\nu'^3n(\nu')/c^2$ is the radiation intensity,
$a>2$, $b>0$. The frequency of the photon decreases in the
course of stimulated scattering in the isotropic medium
therefore the right-hand side of Eq.(\ref{eq_melrose}) is
positive for $\nu'<\nu'_0$. Upon integrating one gets
 \eqb
\frac 1{n(\nu')}\frac{\partial n(\nu')}{\partial t'}= \frac
3{8}\frac{\sigma_TN'cI_0}{m_e\gamma_T^3\nu'_0\nu'}
\Phi\left(\frac{\nu'_0}{4\gamma_T^2\nu'}\right);
 \eqe $$
\Phi(x)=\frac{a+b}{(a-1)(b+1)}e^{-\sqrt{x}}+\frac{x^{-(a-1)}}{a-1}\int_0^{\sqrt{x}}y^{2(a-1)}e^{-y}dy
$$\eqb
-\frac{x^{b+1}}{b+1}\int_{\sqrt{x}}^{\infty}y^{-2(b+1)}e^{-y}dy
=\left\{\begin{array}{ll}(\frac{a+b}{(a-1)(b+1)} & x\ll 1;
\\ \frac{\Gamma(2a-1)}{a-1}x^{-(a-1)}; &
x\gg 1;\end{array}\right.
 \eqe
where $\Gamma(x)$ is the gamma-function. One sees that the
scattering rate is maximal at $\nu'\sim\nu'_0/(2\gamma_T)^2$, the
exact value depending on $a$ and $b$. Substituting $\nu'=\nu'_0/(2\gamma_T)^2$ and $\Phi=1$,
one gets an estimate of the induced scattering rate:
\eqb
\frac 1{n}\frac{\partial n}{\partial t'}= \frac
32\frac{\sigma_TN'cI_0}{m_e\gamma_T\nu'^2_0}.
 \label{rate}\eqe

In order to check the observability condition (\ref{ind_cond}),
one should substitute $\nu'_0=\nu_{\rm obs}/\Gamma$ into the
formula (\ref{rate}) for the induced scattering rate and
express $I_0$  via the observed flux. If the source size is
small so that the proper light travel time, $l'/c$, is less
than the proper expansion time, $r/(c\Gamma)$, the source
radiates within the angle $1/\Gamma$ and the luminosity may be
expressed via the observed flux as $L=(\pi/2)F_{\rm
obs}\nu_{\rm obs}D^2\Gamma^{-4}$. On the other hand, the
luminosity, which is the relativistic invariant, could be
calculated in the proper frame as $L=8\pi^2l'^2\nu'_0I_0$
(assuming the source is spherical in the comoving frame). This
yields
 \eqb
I_0=\frac{F_{\rm obs}D^2}{16\pi\Gamma^3 l'^2};\qquad l'<\frac
r{\Gamma}.
 \eqe
In the opposite case $l'>r/\Gamma$, one can imagine a radially
expanding plasma radiating forward so that the local radiation
flux is $F\nu=4W'c\Gamma^2$, where $W'=4\pi I_0\nu'_0/c$ is the
radiation density in the comoving frame. Then one can write
 \eqb
I_0=\frac{F_{\rm obs}D^2}{16\pi\Gamma r^2};\qquad l'>\frac
r{\Gamma}.
 \eqe
Now the observability condition (\ref{ind_cond}) is written as
 \eqb
\frac{3\sigma_TN'F_{\rm obs}D^2}{32\pi\gamma_Tm_e\nu_{\rm
obs}^2}<\left\{\begin{array}{ll} \Gamma l'; & l'<r/\Gamma;
\\ r; & l'>r/\Gamma.\end{array}\right.
 \label{observ_rel}\eqe
For any specific radiation model, one can check the
observability condition substituting parameters of the emitting
plasma in Eq. (\ref{observ_rel}).

For a rather general preliminary estimate, one can express the
plasma density in the source via the fraction $\zeta$ of the
plasma energy radiated in the pulse. Only a small fraction of
the plasma energy could typically be radiated in the radio band
so that one can expect $\zeta\ll 1$ however one can not exclude
a priori a larger $\zeta$ (and even $\zeta>1$ for a Poynting
dominated source). We will see that this uncertainty is
compensated by a very weak dependence of the result on $\zeta$.
If the source is small, $l'<r/\Gamma$, the total radiated
energy is estimated as $ E_{\rm rad}=\pi D^2\nu_{\rm obs}
F_{\rm obs}\Delta t_{\rm obs}/\Gamma^2$ whereas the total
plasma energy in the source is $E_{\rm pl}=4\pi
l'^3N'\gamma_Tmc^2\Gamma$, where $m=m_e$ in the
electron-positron plasma and $m=m_p$ in the electron-ion
plasma. In the opposite limit, $l'>r/\Gamma$, one should
compare the plasma energy density, $\varepsilon_{\rm
pl}=3mc^3N'\gamma_T\Gamma^2$, with the radiation energy
density, $\varepsilon_{\rm rad}=F_{\rm obs}\nu_{\rm
obs}(D/r)^2$. Now one can write
 \eqb
N'=\frac{F_{\rm obs}\nu_{\rm obs}D^2}{mc^2\gamma_T\zeta}
\left\{\begin{array}{ll} \Delta t/(4l'^3\Gamma^3); &
l'<r/\Gamma;
\\ 1/(3cr^2\Gamma^2); & l'>r/\Gamma.\end{array}\right.
 \eqe
Then the observability condition is written as
 \eqb
\frac{\sigma_TF_{\rm obs}^2D^4}{\zeta\gamma_T^2m_emc^2\nu_{\rm
obs}\Gamma^2}<\left\{\begin{array}{ll} 128\pi l'^4/(3c\Delta
t); & l'<r/\Gamma;
\\ 32\pi r^3; & l'>r/\Gamma.\end{array}\right.
 \eqe
The light travel time arguments imply that $l'<c\Delta t\Gamma$
if $l'<r/\Gamma$ and $r<c\Delta t\Gamma^2$ in the opposite
limit. Taking this into account, one finally finds that the
observability condition implies a lower limit on the Lorentz
factor of the source:
 \eqb
\Gamma>100\frac{F_{\rm
obs,Jy}^{1/4}D^{1/2}_8}{\zeta^{1/8}\gamma_T^{1/4}(m/m_e)^{1/8}(\Delta
t_{\rm s})^{3/8}\nu_{\rm GHz}^{1/8}}.
 \label{Lorents}\eqe

Note that as the radiation within the source is nearly
isotropic, the induced scattering is important only if it
affects the spectrum of the radiation. In the case of the
induced Compton scattering in the relativistically hot plasma,
the photon frequency decreases $\sim 4\gamma_T^2$ times already
in a single scattering therefore the observability condition
for the induced Compton scattering is the condition that the
source is just transparent with respect to this process. The
frequency change in the Raman scattering is small therefore the
corresponding observability condition is less restrictive.

\section{Implications for the observed short extragalactic pulse}

Let us apply the obtained general observability conditions to
the enigmatic radio pulse recently found by Lorimer et al.
(2007) in a pulsar survey at the frequency 1.4 GHz. The
duration of the pulse was $\Delta t\le 5$ ms, the energy in the
pulse $F_{\rm obs}\Delta t=0.15\pm 0.05$ Jy$\cdot$s. The
dispersion measure is an order of magnitude larger than the
expected contribution from the Milky Way and moreover, no
galaxy was found at the position of the source. This lead
Lorimer et al. to conclude that the source of the pulse is on
cosmological distance; they give a very rough estimate $D\sim
500$ Mpc. Origin of this pulse is obscure; Popov \& Postnov
(2007) argue, on the statistical grounds, that this event could
be related to a hyperflare from an extragalactic soft gamma-ray
repeater.

Substituting the parameters of the pulse into Eqs. (\ref{windC}),
one concludes that if the pulse was generated within a stellar
wind, the induced Compton  scattering places the lower limit on
the emission radius, $r>6\cdot 10^{14}(\dot M_{-5}/V_3)^{1/4}$ cm.
A stronger limit is imposed by the Raman scattering; Eq.
(\ref{windR}) yields
 \eqb
 r>5\cdot 10^{15}\left(\frac{\dot M_{-5}}{V_3}\right)^{1/6}\left(\frac D{500 \rm Mpc}\right)^{2/3}
  \rm cm.
 \label{rad}\eqe
 One should note that according to Eq.(\ref{maxangle}), the angle of the Raman
scattering is small in this case, $\theta_{\rm max}=0.024(V_3/\dot
M_{-5})^{1/6}T_6^{-1/2}$, so that the Raman scattering does not
hinder propagation of the pulse. However, scattering even by this
small angle implies the temporal smearing of the pulse above the
observed limit unless the pulse was initially collimated within
the angle $\vartheta<2.4\cdot 10^{-4}(\Delta t/5 {\rm ms})^{1/2}$.
In the last case, the initial radiation flux in the pulse should
have been $(\vartheta/\theta_{\max})^2=10^4\Delta t T_6(\dot
M_{-5}/V_3)^{1/3}$ times larger than that estimated above under
the no scattering assumption; then the induced scattering would
imply even stronger constraints. Therefore in any case, the lower
limit (\ref{rad}) for the emission radius is robust provided the
pulse was generated within the relic stellar wind of the
progenitor star.

The observed pulse was definitely generated within relativistic
plasma. The induced Compton scattering within the source place a
limit on the Lorentz factor of the emitting plasma. Substituting
the observed parameters of the pulse into Eq. (\ref{Lorents}), one
gets
 \eqb
 \Gamma>3800\gamma_T^{-1/4}\left(\frac{\Delta t}{5 \rm ms}\right)^{-5/8}
 \left(\frac{m_e}{\zeta m}\right)^{1/8}\left(\frac D{500 \rm Mpc}\right)^{1/2}.
 \eqe

\section{Conclusions}

In this paper, we have analyzed the effect of induced Compton
and Raman scattering on propagation of a short bright radio
pulse. The work was motivated by predictions that such pulses
could accompany gamma-ray bursts (Usov \& Katz 2000; Sagiv \&
Waxman 2002; Moortgat \& Kuijpers 2005)  and by a recent
discovery of a single extragalactic radio pulse (Lorimer et al.
2007). Macquart (2007) claimed that induced scattering in the
interstellar medium strongly limits the observability of high
brightness temperature transients.
However he ignored two fundamental properties of the process.
First of all, the induced scattering occurs only if the
scattered ray remains within the zone illuminated by the
scattering radiation. In the case of a single short pulse, the
effective optical depth  is determined by the duration of the
pulse but not by the scale of the scattering medium. Therefore
a short enough pulse could propagate freely through the
interstellar medium. The second important property is that in
the presence of a powerful radiation beam, even a weak
background radiation grows exponentially via the induced
scattering of the beam photons. If the primary beam is narrow,
the scattering outside the beam dominates because the
scattering within the beam is suppressed by the recoil factor
$1-\mathbf{\Omega\cdot\Omega}_1$ in the scattering rate.
Outside the source, the radiation subtends a small solid angle
therefore the induced scattering in the surrounding medium
occurs outside the beam. In this case, the effective optical
depth depends not on the brightness temperature and the angle
subtended by the primary radiation, which could not be found
separately without model assumptions, but only on the radiation
flux, which is a directly observable quantity. We demonstrated
that the induced scattering in the surrounding medium could
hinder the escape of a bright short pulse only if the source is
embedded into a dense medium, like stellar wind. We estimated a
limiting radius beyond which the pulse could propagate.

One should stress that these estimates assume a single pulse.
If a sequence of pulses (e.g., pulsar emission) propagates in
the medium with the characteristic scale exceeding the distance
between the pulses, the induced scattering occurs as if the
emission was steady with the average radiation flux. On the
other hand, our results could be applied to giant pulses from
pulsars, which are rare enough to be considered as isolated
phenomena.

We have also analyzed induced scattering within the
relativistically moving source. Transparency of the source is
determined by the radiation intensity and by amount of plasma
within the source. 
Introducing a fraction $\zeta$ of the plasma energy emitted in
the pulse, we found a lower limit on the Lorentz factor of the
source, which turned out to be very weakly dependent on
$\zeta$.

\acknowledgements
This work was supported by the German-Israeli Foundation for
Scientific Research and Development.

\clearpage
\begin{figure*}
\includegraphics[scale=0.6]{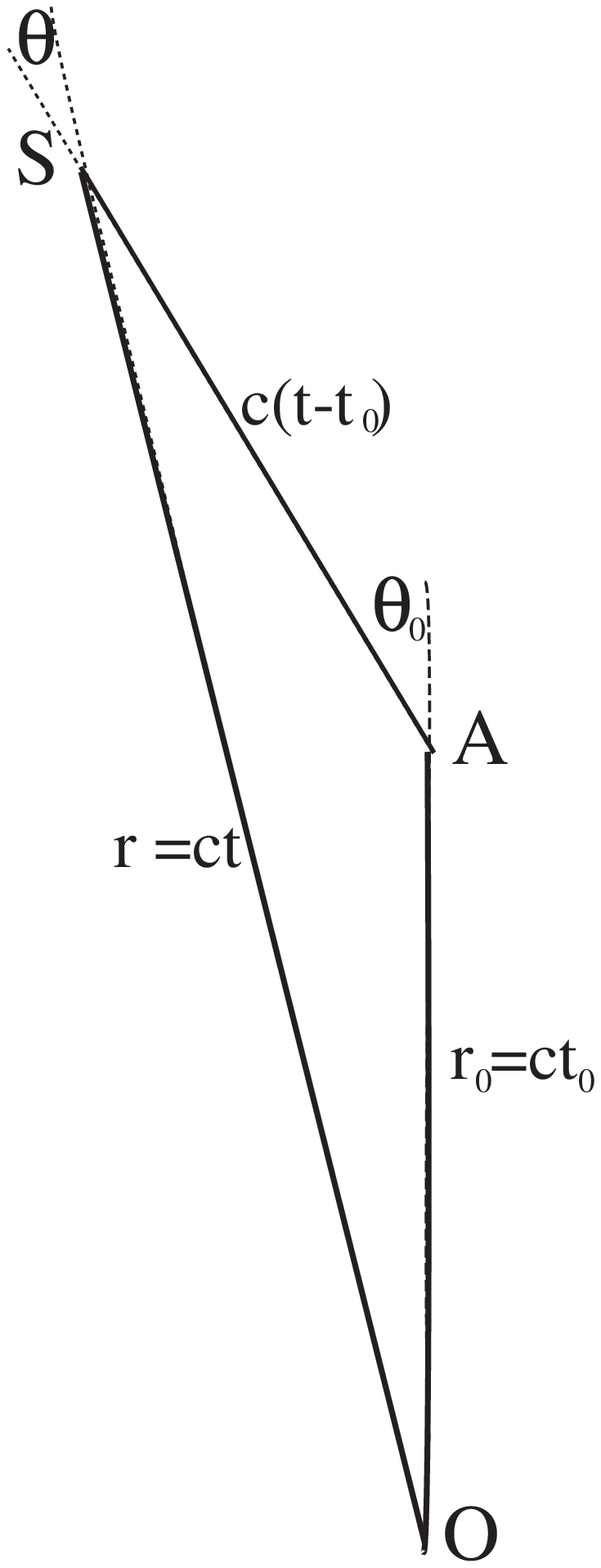}
\caption{Geometry of the scattering. The primary pulse propagates
radially from the point O. Just when it arrives at the point A (at
the time $t_0$), a seed ray is launched at the angle $\theta_0$.
At the time $t$, the seed ray arrives at the point S where it
makes the angle $\theta$ with the radial direction.}
\end{figure*}
\end{document}